\documentclass[twocolumn,superscriptaddress,showpacs,preprintnumbers,amsmath,amssymb]{revtex4}

\def\lesssim{\ \raise.3ex\hbox{$<$}\kern-0.8em\lower.7ex\hbox{$\sim$}\ }
\def\gesim{\ \raise.3ex\hbox{$>$}\kern-0.8em\lower.7ex\hbox{$\sim$}\ }

\usepackage{graphicx}
\usepackage{dcolumn}
\usepackage{bm}
\usepackage{amsmath,amssymb}

\begin{document}

\title{Photoemission spectrum and effect of inhomogeneous pairing fluctuations in the BCS-BEC crossover regime of an ultracold Fermi gas}

\author{Shunji Tsuchiya}
\affiliation{Department of Physics, Keio University, 3-14-1 Hiyoshi, Kohoku-ku, Yokohama 223-8522, Japan}
\affiliation{CREST(JST), 4-1-8 Honcho, Saitama 332-0012, Japan}
\author{Ryota Watanabe}
\affiliation{Department of Physics, Keio University, 3-14-1 Hiyoshi, Kohoku-ku, Yokohama 223-8522, Japan}
\author{Yoji Ohashi}%
\affiliation{Department of Physics, Keio University, 3-14-1 Hiyoshi, Kohoku-ku, Yokohama 223-8522, Japan}
\affiliation{CREST(JST), 4-1-8 Honcho, Saitama 332-0012, Japan}

\date{\today}

\begin{abstract}
We investigate the photoemission-type spectrum in a cold Fermi gas which
 was recently measured by JILA group [J. T. Stewart {\it et al}., Nature
 \textbf{454}, 744 (2008)]. This quantity gives us very useful
 information about single-particle properties in the BCS-BEC
 crossover. In this letter, including pairing fluctuations within a
 $T$-matrix theory, as well as effects of a harmonic trap within the
 local density approximation, we show that spatially inhomogeneous
 pairing fluctuations due to the trap potential is an important key to
 understand the observed spectrum. In the crossover region, while strong
 pairing fluctuations lead to the so-called pseudogap phenomenon in the
 trap center, such strong-coupling effects are found to be weak around
 the edge of the gas. Our results including this effect are shown to
 agree well with the recent photoemission data by JILA group. 

\end{abstract}

\maketitle

\par
The recent photoemission-type experiment developed by JILA
group\cite{Stewart} provides a powerful method to study microscopic properties
of cold Fermi gases. This experiment is an analogue of angle-resolved
photoemission spectroscopy (ARPES)\cite{Damascelli}, which has been
extensively applied in condensed matter physics. Using this technique, one
can probe single-particle excitations that allow us to investigate
many-body effects in the BCS-BEC crossover\cite{reviews,Regal,Zwierlein,Kinast,Bartenstein,Eagles,Leggett,NSR,SadeMelo,Timmermans,Holland,Ohashi}. 
Indeed, the observed spectra exhibit dramatic change in the crossover region\cite{Stewart}.
\par
One typical many-body effect on single-particle excitations expected in
the BCS-BEC crossover regime of a cold Fermi gas is the pseudogap
effect. In this phenomenon, preformed pairs cause a gap-like structure
in the density of states (DOS) even above the superfluid phase
transition temperature $T_{\rm c}$. The pseudogap has been observed in
the underdoped regime of high-$T_{\rm c}$ cuprates\cite{Lee}. However,
as the origin of this phenomenon in high-$T_{\rm c}$ cuprates, in
addition to pairing fluctuations, various possibilities have been
proposed, such as spin fluctuations and a hidden order. Since the
BCS-BEC crossover in a cold Fermi gas is dominated by pairing
fluctuations, the study of pseudogap in this system would be also useful
for clarifying the validity of the pseudogap mechanism based on
preformed pairs in high-$T_{\rm c}$ cuprates. Since the pseudogap
appears in single-particle excitations, the photoemission experiment
would be useful for this purpose.
\par
In considering the pseudogap effect in cold Fermi gases, one should note
that the presence of a trap naturally leads to spatially inhomogeneous
pairing fluctuations. While the pseudogap structure in DOS is expected
to be remarkable in the trap center, such a many-body effect may be weak
around the edge of gas cloud. Since the pseudogap in high-$T_{\rm c}$
cuprates is a uniform phenomenon, this inhomogeneous pseudogap effect is
unique to trapped Fermi gases. Indeed, the double-peak structure in the
photoemission spectrum observed in the strong-coupling BEC regime of
$^{40}$K Fermi gas\cite{Stewart} seems difficult to explain as far as a
simple uniform system is considered\cite{Tsuchiya}. 
\par
In this paper, we study pseudogap phenomena in a trapped Fermi gas above
$T_{\rm c}$, addressing the recent photoemission experiment by JILA
group\cite{Stewart}. Extending our previous paper for a uniform Fermi
gas to include effects of a harmonic trap within the local density
approximation (LDA), we calculate the local DOS, as well as the local
spectral weight (SW), over the entire BCS-BEC crossover region within
the $T$-matrix approximation in terms of pairing
fluctuations\cite{Janko,Rohe,Yanase,Perali,Tsuchiya}. 
We clarify how the inhomogeneous pseudogap phenomenon appears in these
quantities. 
\par
Recently, Refs.~\cite{Chen,Dao} have studied the photoemission spectra using
phenomenological theories with the BCS ansatz. While they capture some
features of experimental results in the unitarity limit, they cannot
explain the observed spectrum in the BEC regime, which consists of upper
sharp and lower broad peaks. In contrast, including the inhomogeneous
strong-coupling effect, we show that our {\it ab initio} calculation of the
photoemission spectra can naturally explain the experiments\cite{Stewart} from the
weak-coupling BCS to the strong-coupling BEC regimes, without
introducing any free parameters.
\par
We consider a two-component Fermi gas in a harmonic trap. Assuming a
broad Feshbach resonance, we employ the ordinary BCS model, described by
the Hamiltonian,
\begin{eqnarray}
H&=&\sum_{\bm p,\sigma}\xi_{\bm p} c_{\bm p\sigma}^\dagger c_{\bm p\sigma}
\nonumber
\\
&-&
U\sum_{\bm q,\bm p,\bm p^\prime}c_{\bm p+\bm q/2\uparrow}^\dagger c_{-\bm p+\bm q/2\downarrow}^\dagger c_{\bm p^\prime+\bm q/2\downarrow}c_{\bm p^\prime+\bm q/2\uparrow}.
\end{eqnarray}
Here, $c_{\bm p\sigma}$ is the annihilation operator of a Fermi atom
with pseudospin $\sigma=\uparrow,\downarrow$, and the kinetic energy
$\xi_{\bm p}=\varepsilon_{\bm p}-\mu$ is measured from the chemical
potential $\mu$, where $\varepsilon_{\bm p}=p^2/2m$ and $m$ is the
atomic mass. The pairing interaction $-U(<0)$ is assumed to be tunable
by a Feshbach resonance\cite{Chin}, which is related to the $s$-wave
scattering length $a_s$ as\cite{Randeria} $4\pi a_s/m=-U/[1-U\sum_{\bm
p}1/(2\varepsilon_{\bm p})]$.
\par
Within a LDA, effects of a trap are conveniently incorporated into the
theory by replacing the chemical potential $\mu$ by $\mu(r)\equiv
\mu-V(r)$\cite{Pethick}, where $V(r)=m\omega_{\rm tr}^2r^2/2$ is a
harmonic potential (where $\omega_{\rm tr}$ is a trap frequency). The
LDA single-particle Green's function is given by $G_{\bm
p}(i\omega_n,r)=1/[i\omega_n-\xi_{\bm p}(r)-\Sigma_{\bm
p}(i\omega_n,r)]$, where $\xi_{\bm p}(r)=\varepsilon_{\bm p}-\mu(r)$,
and $\omega_n$ is the fermion Matsubara frequency. The LDA self-energy
$\Sigma_{\bm p}(i\omega_n,r)$ involves effects of pairing fluctuations
within the $T$-matrix
approximation\cite{Janko,Rohe,Yanase,Perali,Tsuchiya}, which is given by
$\Sigma_{\bm p}(i\omega_n,r)=T\sum_{\bm q,i\nu_n}\Gamma_{\bm
q}(i\nu_n,r)G^0_{\bm q-\bm p}(i\nu_n-i\omega_n,r)$. Here, $\nu_n$ is the
boson Matsubara frequency, and $G_{\bm
p}^0(i\omega_n,r)=1/[i\omega_n-\xi_{\bm p}(r)]$ is the LDA free fermion
propagator. $\Gamma_{\bm q}(i\nu_n,r)=-U/[1-U\Pi_{\bm q}(i\nu_n,r)]$ is
the particle-particle scattering matrix within the $T$-matrix
approximation, where $\Pi_{\bm q}(i\nu_n,r)=T\sum_{\bm
p,i\omega_n}G_{\bm p+\bm q/2}^0(i\nu_n+i\omega_n,r)G^0_{-\bm p+\bm
q/2}(-i\omega_n,r)$ is the pair propagator\cite{NSR}.
\par
In this paper, we focus on the case at $T_{\rm c}$ to compare with the
experiments in Ref.~\cite{Stewart}. The region above $T_{\rm c}$ will be
discussed elsewhere. In LDA, the superfluid phase transition is
determined by the Thouless criterion at the trap center\cite{Ohashi2},
i.e., $\Gamma_{\bm q=0}(i\nu_n=0,r=0)^{-1}=0$\cite{NSR}. Strong-coupling
effects on $\mu$ are included by solving this $T_{\rm c}$-equation
together with the equation for the total number of Fermi atoms,
$N=2T\int d{\bm r}\sum_{\bm p,i\omega_n}G_{\bm
p}(i\omega_n,r)e^{i\omega_n\delta}$. 
\par
\begin{figure}
\centerline{\includegraphics[width=6cm]{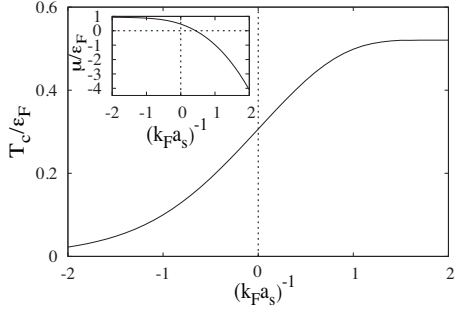}}
\caption{Calculated $T_{\rm c}$ as a function of the inverse scattering
 $a_s^{-1}$. The inset shows $\mu$ at $T_{\rm c}$. $k_{\rm F}$ is the
 Fermi momentum, and $\varepsilon_{\rm F}$ is the Fermi energy. We use
 these $T_{\rm c}$ and $\mu$ in Figs.~\ref{swspat}-\ref{dosave}.} 
\label{fig1}
\end{figure}

Figure~\ref{fig1} shows the calculated $T_{\rm c}$ and $\mu$ as
functions of $(k_Fa_s)^{-1}$, where $k_F$ is the Fermi momentum
defined from the Fermi energy $\varepsilon_F=(3N)^{1/3}\omega_{\rm tr}=k_F^2/(2m)$. 
In Fig.~\ref{fig1}, $T_{\rm c}$ in a trapped gas is found to be higher than
that in a uniform gas in terms of $\varepsilon_F$\cite{Ohashi2,Perali2}. Using them, we calculate
the local SW $A(\bm p,\omega,r)$ and local DOS $\rho(\omega,r)$ from the
analytic continued Green's function as, respectively,
\begin{equation}
A(\bm p,\omega,r)=-\frac{1}{\pi}{\rm Im}[G_{\bm p}(i\omega_n\to\omega+i\delta,r)],
\end{equation}
\begin{equation}
\rho(\omega,r)=\sum_{\bm p}A(\bm p,\omega,r).
\end{equation}

\par
The photoemission spectrum\cite{Stewart} can be calculated in the same
way as the rf-tunneling current spectroscopy\cite{Torma,He,Ohashi2},
where atoms in one of the two pseudospin states
($\equiv|\uparrow\rangle$) are outcoupled into an unoccupied state
$|3\rangle$ ($\ne |\uparrow,\downarrow\rangle$) by an applied rf
pulse. Since the final-state interaction can be safely neglected in
$^{40}$K Fermi gas\cite{Stewart,Chen}, the Hamiltonian for $|3\rangle$
is simply given by $H_3=\sum_{\bm
p}[\varepsilon_{\bm p}+\omega_3-\mu_3(r)]b_{\bm p}^\dagger b_{\bm p}$. Here,
$b_{\bm p}$ describes the third pseudospin state $|3\rangle$, and
$\omega_3$ is the energy difference between $|\uparrow\rangle$ and
$|3\rangle$. The chemical potential $\mu_3(r)=\mu_3-V(r)$ involves trap
effects within LDA. The transition from $|\uparrow\rangle$ to
$|3\rangle$ is induced by the tunneling
Hamiltonian\cite{Torma,He,Ohashi2} $H_T=t_{\rm F}\sum_{\bm
k}[e^{-i\omega_L t}b_{\bm k+\bm q_L}^\dagger c_{\bm k\uparrow}+{\rm
H.c.}]$, where $t_F$ is a transfer matrix element, and ${\bm q}_L$ and
$\omega_L$ are the momentum and frequency of the rf-pulse,
respectively. 
The photoemission spectrum is obtained from rf-tunneling current
$I(\Omega,r)$ from $|\uparrow\rangle$ to $|3\rangle$, where
$\Omega\equiv\omega_L-\omega_3$ is the rf-detuning. 
Within the linear response theory, we obtain
$I(\Omega,r)=-i\int_{-\infty}^{t}dt^\prime~\langle[\hat
J(r,t),H_T(r,t^\prime)]\rangle e^{\delta t^\prime}$, where $J(t)\equiv
-it_{\rm F}\sum_{\bm k}e^{i(H+H_3) t}[e^{-i\omega_Lt}b_{\bm k+\bm
q_L}^\dagger c_{\bm k\uparrow}-{\rm H.c.}]e^{-i(H+H_3) t}$ is the
tunneling current operator in the Heisenberg representation, and
$H_T(t)\equiv e^{i(H+H_3)t}H_T e^{-i(H+H_3) t}$. 
We thus obtain the rf-tunneling current $I(\Omega,r)=\sum_{\bm p}I(\bm
p,\Omega,r)$, where the momentum-resolved photoemission spectrum $I(\bm
p,\Omega,r)$ has the form
\begin{equation}
I(\bm p,\Omega,r)=2\pi t_F^2 A(\bm p,\xi_{\bm p}(r)-\Omega,r)f(\xi_{\bm p}(r)-\Omega).
\label{rfcurrent}
\end{equation}
Here, $f(\Omega)$ is the Fermi distribution function. In
Eq.~(\ref{rfcurrent}), we have assumed that the momentum $\bm q_L$ of
rf-photon is negligible and $|3\rangle$ is initially empty ($f(\varepsilon_p-\mu_3(r))=0$).
\par
\begin{figure}
\centerline{\includegraphics[width=6.5cm]{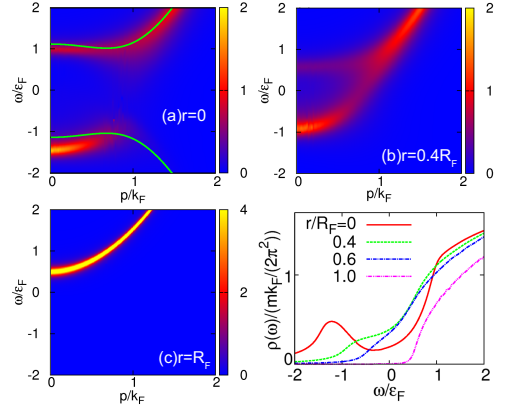}}
\caption{(color online). (a)-(c) Intensity of local SW $A(\bm p,\omega,r)$ (units of
 $\varepsilon_F^{-1}$). (d) local DOS $\rho(\omega,r)$. We set $T=T_{\rm 
 c}$ and $(k_{\rm F}a_s)^{-1}=0$. The dotted line in (a) is the BCS-like
 quasiparticle spectrum with $\Delta_{\rm pg}=1.06$.} 
\label{swspat}
\end{figure}

In a uniform Fermi gas, the photoemission spectrum is related to SW of
occupied states as $I(\bm p,\Omega\to \xi_{\bm p}-\omega)=2\pi t_F^2A(\bm
p,\omega)f(\omega)$. In particular, it is equal to SW below $\omega=0$
at $T=0$.  When $T>0$, thermally excited quasiparticles also contribute
to the spectrum, so that $I(\bm p,\Omega\to \xi_{\bm p}-\omega)$ becomes
finite even when $\omega>0$.  
\par
In the photoemission experiment\cite{Stewart}, since the rf-pulse is applied to the whole gas cloud, the observed spectrum involves contributions from all spatial regions of the cloud. To include this situation, we should take spatial average of Eq.(\ref{rfcurrent}) as
\begin{equation}
I_{\rm ave}(\bm p,\Omega)=\frac{2\pi t_F^2}{V}\int d\bm r\  A(\bm
p,\xi_{\bm p}(r)-\Omega,r)f(\xi_{\bm p}(r)-\Omega).  
\label{curave}
\end{equation}
Here, $V=4\pi R_F^3/3$, where $R_F=\sqrt{2\mu/(m\omega_{\rm tr}^2)}$ is
the Thomas-Fermi radius\cite{note}. We emphasize that 
this gives a proper definition for spatially averaged photoemission
spectrum. For later convenience, we define the averaged occupied SW and DOS by,
respectively,
\begin{equation}
\overline{A(\bm p,\omega)f(\omega)}\equiv I_{\rm ave}(\bm p,\Omega\to\xi_p-\omega)/(2\pi t_F^2),
\label{eq.sw}
\end{equation}
\begin{equation}
\overline{\rho(\omega)f(\omega)}\equiv\sum_{\bm p}I_{\rm ave}(\bm p,\Omega\to\xi_p-\omega)/(2\pi t_F^2).
\label{eq.dos}
\end{equation}
\par

\begin{figure}
\centerline{\includegraphics[width=6.5cm]{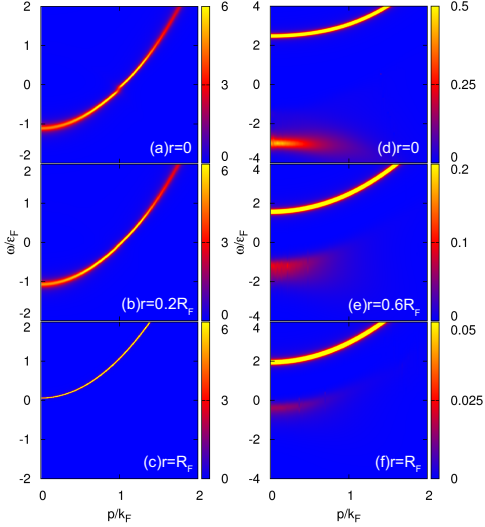}}
\caption{(color online). Intensity of local SW $A(\bm p,\omega,r)$ (units
 of $\varepsilon_F^{-1}$). 
(a)-(c) $(k_{\rm F}a_s)^{-1}=-1$. (d)-(f) $(k_{\rm F}a_s)^{-1}=+1$. We
 set $T=T_{\rm c}$.}
\label{fig3}
\end{figure}

To see the basic characters of Eqs.~(\ref{eq.sw}) and (\ref{eq.dos}), it
is helpful to consider a free Fermi gas at $T=0$. In this case,
Eqs. (\ref{eq.sw}) and (\ref{eq.dos}) reduce to, respectively, 
\begin{equation}
\overline{A(p,\omega)f(\omega)}=|\omega/\mu|^{3/2}\delta(\omega-\xi_p)\theta(-\omega),
\label{eq.freesw}
\end{equation}
\begin{equation}
\overline{\rho(\omega)f(\omega)}=(m^{3/2}/\sqrt{2}\pi^2)|\omega/\mu|^{3/2}\sqrt{\omega+\mu}\theta(\omega+\mu)\theta(-\omega). 
\label{eq.freedos}
\end{equation}
In the former, the peak position gives the one-particle energy $\xi_{\bm p}$. In the latter, DOS in a uniform gas ($\propto\sqrt{\omega+\mu}$) is modified by the factor $|\omega|^{3/2}$.
\par
We now show our numerical results. Figure~\ref{swspat} shows the local
SW $A({\bm p},\omega,r)$ and DOS $\rho(\omega,r)$ at $T_{\rm c}$ in the
unitarity limit ($(k_Fa_s)^{-1}=0$). In the trap center (panel (a)), a
clear pseudogap structure exists, i.e., two prominent peaks appear along
the particle branch and hole branch of the BCS-like quasiparticle
spectrum $\omega=\pm\sqrt{\xi_{\bm p}^2+\Delta_{\rm pg}^2}$, where the
superfluid gap $\Delta$ is replaced by the pseudogap $\Delta_{\rm
pg}$. The origin of $\Delta_{\rm pg}$ is a particle-hole coupling by
pairing fluctuations\cite{Janko,Rohe,Yanase,Perali,Tsuchiya}. Since
pairing fluctuations also induce a finite lifetime of quasiparticle
excitations, the particle and hole branches in Fig.~\ref{swspat}(a) have
finite widths, which is in contrast to the mean-field BCS case, where
both branches appear as $\delta$-functional peaks.
We note that the deviation of the lower peak from the BCS-like quasiparticle spectrum
is considered due to the presence of excited pairs \cite{Tsuchiya,Perali}.
\par
Pairing fluctuations are weak around the edge of the gas due to low
particle density. Thus, the pseudogap in the local SW gradually
disappears, as one leaves from the trap center. (See
Figs. \ref{swspat}(b) and (c).) In panel (c), a single sharp peak line
only exists near the one-particle energy of a free Fermi gas $\xi_{\bm
p}(R_{\rm F})$. 
Namely, the gas is pseudogapped in the center of the trap, while it is
nearly non-interacting on its edges.
\par
These inhomogeneous features can be also seen in local DOS. As shown in
Fig.~\ref{swspat}(d), while a large dip structure (which is a
characteristic pseudogap effect in DOS) appears around $\omega=0$ in
the trap center, local DOS is almost equal to that of a free Fermi gas
($\rho(\omega,r)=(m^{3/2}/(\sqrt{2}\pi^2))\sqrt{\omega+\mu(r)}$) when
$r=R_{\rm F}$. Since spatial inhomogeneity is unique to trapped Fermi
gases, the observation of local SW and DOS by using the tomographic
techniques\cite{Shin,Schirotzek} would be interesting.
\par
\begin{figure}
\centerline{\includegraphics[width=7cm]{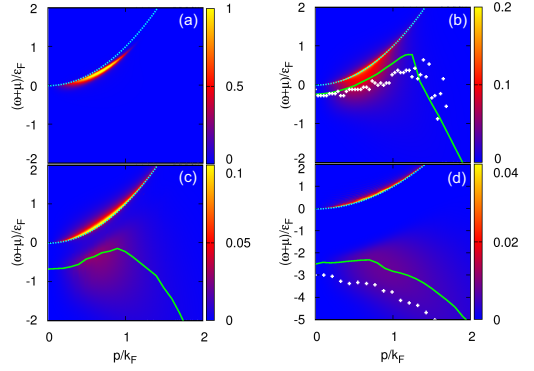}}
\caption{(color online). Intensity of averaged occupied SW $p^2\overline{A(\bm
p,\omega)f(\omega)}$ (units of $1/2m$) at $T_{\rm c}$. The values of
 pairing interaction $(k_Fa_s)^{-1}$ are (a) -1, (b) 0, (c) 0.4, and (d)
 1. The dashed line indicates the spectrum of a free Fermi gas
 $\xi_p$. The solid line is the lower peak position of intensity. Symbols are
 corresponding experimental data \cite{Stewart}.
} 
\label{pesave}
\end{figure}

We note that the inhomogeneous pseudogap structure depends on the
strength of pairing interaction. When pairing fluctuations are weak in
the BCS regime, the double-peak structure in
$A(\bm p,\omega,r)$ soon disappears, as one leaves from
the trap center. (See Figs.~\ref{fig3}(a)-(c).) In contrast,
Figs.~\ref{fig3}(d)-(f) show that the pseudogap features persist even
near the edge of the gas in the BEC regime.
\par

\begin{figure}
\centerline{\includegraphics[width=5.5cm]{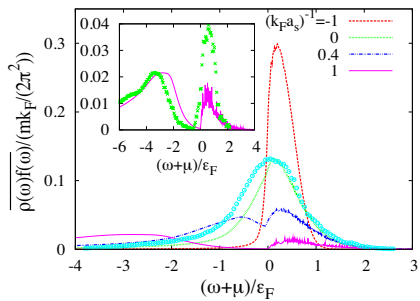}}
\caption{(color online). Averaged occupied DOS $\overline{\rho(\omega)f(\omega)}$ at
 $T_{\rm c}$. 
The inset shows a magnification of the curve for $(k_Fa_s)^{-1}=1$.
Circles and crosses show the experimental data for $(k_Fa_s)^{-1}=0$
 and $1$, respectively\cite{Stewart}. The data are fitted so that the peak value
 coincides with that of theory curves. The data agree well with the theory curves.
}
\label{dosave}
\end{figure}

Figures \ref{pesave} and \ref{dosave}, respectively, show the averaged
occupied SW and DOS at $T_{\rm c}$. In the weak-coupling BCS regime,
these quantities are expected to be close to the free Fermi gas results
in Eqs.~(\ref{eq.freesw}) and (\ref{eq.freedos}). Indeed, the position of
the peak line in Fig. \ref{pesave}(a) is almost at the single-particle
energy of a free fermion $\omega=\xi_{\bm p}$ ($\le 0$), and the
averaged occupied DOS in Fig.~\ref{dosave} is close to DOS for a free
Fermi gas multiplied by $|\omega|^{3/2}$ when $(k_{\rm
F}a_s)^{-1}=-1$. However, we also find that the peak line in
Fig.~\ref{pesave}(a) is slightly below the curve of $\omega=\xi_p$,
which is a signature of pseudogap effect near the trap center. Namely,
the pseudogap $\Delta_{\rm pg}$ lowers the hole branch as $-|\xi_{\bm
p}(r)|\to -\sqrt{\xi_{\bm p}^2(r)+\Delta_{\rm pg}^2}$, and this effect
still remains even after spatial average.
\par
In the crossover region, pairing fluctuations are strong around the trap
center. In the unitarity limit, while the region around the edge of the
gas still gives a sharp peak line at $\omega\simeq \xi_{\bm p}$ in
$\overline{A(\bm p,\omega)f(\omega)}$, the pseudogap in SW around $r=0$
causes the broadening of the lower part of the peak line around
$p/k_{\rm F}\simeq 0.5$ shown in Fig.~\ref{pesave}(b). In addition,
short lifetime of quasiparticle excitations by strong pairing
fluctuations around the trap center also causes the broadening of
$\overline{\rho(\omega)f(\omega)}$, as shown in Fig. \ref{dosave}.
We also note that one can see the back-bending of the lower peak
position in Figs.~4(b)-(d) originating from the lower branch shown in
Fig.~2(a), which well agrees with the experimental data\cite{Stewart}.
\par
In the BEC regime, the pseudogap features persist to the edge of the
gas. Thus, the double-peak structure in the local SW (See
Figs. \ref{fig3}(d)$\sim$(f).) is not smeared out by spatial average, as
shown in Figs. \ref{pesave}(c)$\sim$(d). In particular, the double-peak
structure in panel (d), consisting of upper {\it sharp} peak and lower
{\it broad} peak, agrees well with the photoemission experiment by JILA
group\cite{Stewart}. 
We emphasize that this spectral structure cannot be explained within the
previous phenomenological theories\cite{Chen,Dao}.
Such a double-peak structure also appears in
Fig. \ref{dosave}, which is also consistent with the
experiment\cite{Stewart}.  
\par
In the BEC limit, the spectral weight only has the upper sharp peak describing the dissociation of two-body bound molecules. On the other hand, as discussed in Ref.~\cite{Tsuchiya}, the lower broad peak in SW in the BEC regime is an evidence of many-body character of paired atoms. Thus, the double-peak structure in Figs. \ref{pesave}(c)$\sim$(d) may be understood as a result of the fact that, in the BCS-BEC crossover, the character of fermion pair continuously changes from the many-body bound state associated with the Cooper instability to the two-body bound state where the Fermi surface is not necessary.

\par
To summarize, we have discussed inhomogeneous pseudogap effect in the BCS-BEC crossover regime of a trapped Fermi gas. Including this effect, we showed that calculated spatially averaged photoemission spectrum agrees well with the recent experiment by JILA group\cite{Stewart}. Since spatial inhomogeneity by a trap is inevitable in a cold Fermi gas, our results would be useful for clarifying strong-coupling phenomena of this system over the entire BCS-BEC crossover region. 
\par
We wish to thank A. Griffin for useful discussions. We acknowledge
J. P. Gaebler and D. S. Jin for providing us with their experimental data.



\begin{thebibliography}{}
\bibitem{note}The spatial integration is taken up to $1.5R_F$, so that the occupied DOS converges.
\end{thebibliography}
\end{document}